\documentclass[12pt,a4paper,notitlepage]{article}

\usepackage{amssymb,amsmath}
\usepackage[dvips]{graphicx}

\newcommand{\bsl}{\boldsymbol}
\newcommand{\rr}{\mathbf{r}}

\begin{document}

\author{Yu.A.Simonov\thanks{e-mail: simonov@itep.ru}~ and
M.A.Trusov\thanks{e-mail: trusov@itep.ru} \\[4mm]
\textit{Institute of Theoretical and Experimental Physics, Moscow,
Russia}}

\title{Axial and tensor charge of the nucleon in the Dirac orbital model}


\maketitle

\begin{abstract}
\noindent Using the expansion of the baryon wave function in a
series of products of single quark bispinors (Dirac orbitals), the
nonsinglet axial and tensor charge of the nucleon are calculated.
The leading term yields ${G_A}/{G_V} = 1.27$ and in good agreement
with experiment. Calculation is essentially parameter-free and
depends on the string tension $\sigma$ and $\alpha_s$, fixed  at
standard values. The importance of lower Dirac bispinor component,
yielding 18\% to the wave function normalization is stressed.
\end{abstract}

\newpage

Axial and tensor charges of nucleons are important to characterize
the basic structure of the nucleon as  composed of strongly
coupled quarks \cite{1}-\cite{3}. It is known that nonrelativistic
quark models predict $\frac{G_A}{G_V}=\frac{5}{3}$ in strong
disagreement with experimental value 1.27, while for massless
relativistic quarks, e.g. in the MIT bag model, one obtains much
smaller values $\frac{G_A}{G_V}=1.09$. Thus the calculation of
$\frac{G_A}{G_V}$  (and tensor charge $\delta  q$) gives a clue to
the relativistic dynamics of quarks in the nucleon.

Moreover, in a recent paper \cite{4} it was shown that the
knowledge of the ratio of  $\frac{G_A}{G_V}$  for baryon decays is
important for the accurate determination of the CKM matrix element
$V_{us}$ These considerations justify the systematic analysis of
baryon decays in the framework of Dirac Orbital Expansion (DOE)
the first part of which is reported below.

The contribution of vector and axial hadronic currents, $V_\mu=
i\bar \psi_u  \gamma_\mu \psi_d$,  $A_\mu = i \bar u\gamma_\mu
\gamma_5 \psi_d$, to the neutron $\beta$-decay is  characterized
by the ratio
\begin{equation}
\frac{G_A}{G_V}=\frac {\bigl\langle p_\uparrow
\bigm|A_z\bigm|n_\uparrow\bigr\rangle}{\bigl\langle p_\uparrow
\bigm|V_0\bigm|n_\uparrow\bigr\rangle}, \label{1}
\end{equation}
where $p_\lambda, n_\lambda, \lambda=\pm \frac{1}{2}$ are proton
and neutron wave functions with spin projection $\lambda $
\cite{1}-\cite{3}. In a similar way the tensor charge is expressed
through the proton matrix element of the tensor current
$T_{\mu\nu}\bar \psi i \sigma_{\mu\nu} \gamma_5 \psi$ \cite{5}.

To construct the baryon wave function, one starts with the
Hamiltonian \cite{6} obtained in the instantaneous approximation
from the general Bethe--Salpeter equation:
\begin{equation}
\Hat{H}\Psi(\rr_1,\rr_2,\rr_3)=E\Psi,\quad \Hat{H}=\sum_{i=1}^3
\Hat{H}_i+\Delta H \label{2}
\end{equation}
with
\begin{equation}
\Hat{H}_i=\mathbf{p}_{(i)}
\bsl{\alpha}_{(i)}+\beta_{(i)}(m_i+M(\rr_i)) \label{3}
\end{equation}
where $M(\rr_i)$ in the limit of vanishing gluon correlation
length is
\[
M=\sigma |\rr_i|e^{i\gamma_5\Hat{\phi}(\rr_i)}
\]
and $\rr_i=\mathbf{x}_i-\mathbf{x}_0$, $\mathbf{x}_0$ is the
string-junction coordinate and $\Hat{\phi}(\rr_i)$ is the
Nambu--Goldstone octet. Here $\Delta H$ contains perturbative
gluon exchanges. We expand the baryon wave function in a series of
products of quark eigenfunctions
$\psi_n^{(i)}=\binom{v^{(i)}}{w^{(i)}}$, namely \cite{7,8}
\begin{equation}
\Psi(\rr_1,\rr_2,\rr_3)=\sum_{\{n_i\}}\prod_{i=1}^3
\psi_{n_i}^{(i)}(\rr_i)C_{n_1n_2n_3} \label{4}
\end{equation}

In what follows we shall consider the leading valence
approximation for the nucleon keeping only the first term in
(\ref{4}), $\Psi\to\Psi_0$, which contains the ground state
$S$--wave Dirac orbitals $|u_\lambda\rangle$ and
$|d_\lambda\rangle$ for $u$ and $d$ quarks with spins up and down.
One has
\begin{gather}
|p\uparrow\rangle=\sqrt{\frac{1}{18}}\left[-2(|u\uparrow u\uparrow
d\downarrow \rangle+\text{\textit{perm.}})+(|u\uparrow u\downarrow
d\uparrow\rangle+\text{\textit{perm.}} )\right] \label{5} \\
|n\uparrow\rangle=\sqrt{\frac{1}{18}}\left[-2(|d\uparrow d\uparrow
u\downarrow \rangle+\text{\textit{perm.}})+(|d\uparrow d\downarrow
u\uparrow\rangle+\text{\textit{perm.}} )\right] \label{6}
\end{gather}
The expressions (\ref{5}) and (\ref{6}) have the same form as  in
the standard $SU(4)$ or $SU(6)$ model \cite{3} except for the
bispinor contents of $|u_\lambda\rangle$ and $|d_\lambda\rangle$.

Insertion of (\ref{5}), (\ref{6}) into (\ref{1})
yields\footnote{The sign of $G_A$ corresponds to \cite{1,a}}
\begin{equation}
\frac{G_A}{G_V}=
+\frac{5}{3}\Bigl\langle\chi_\uparrow\Bigm|\bsl{\Sigma}_3\Bigm|\chi_\uparrow\Bigr\rangle,
\quad \bsl{\Sigma}=\begin{pmatrix} \bsl{\sigma} & 0 \\ 0 &
\bsl{\sigma}
\end{pmatrix} \label{7}
\end{equation}
where $\chi_\uparrow$ is
\begin{equation}
\chi_\uparrow(r,\theta,\phi)=\frac{1}{r}
\binom{G(r)\Omega_{\frac{1}{2},0,\frac{1}{2}}(\theta,\phi)}
{iF(r)\Omega_{\frac{1}{2},1,\frac{1}{2}}(\theta,\phi)};\quad
\int\limits_0^\infty \left[G^2(r)+F^2(r)\right] dr=1 \label{8}
\end{equation}

To take into account perturbative gluon exchange we represent
$\Delta H$ effectively as one-particle operators,
\[
\Delta H=\sum_{i=1}^3 \left(-\frac{\zeta}{r_i}\right)
\]
and the equations for $G(r)$, $F(r)$ acquire the form \cite{b}
\begin{equation}
\label{9}
\begin{gathered}
G'-\frac{1}{r}G-\left(E+m+\sigma r+\frac{\zeta}{r}\right)F=0,\\
F'+\frac{1}{r}F+\left(E-m-\sigma r+\frac{\zeta}{r}\right)G=0
\end{gathered}
\end{equation}
Finally $G_A$ can be written as
\begin{equation}
G_A=+\frac{5}{3}\left(1-\frac{4}{3}\eta\right),\quad
\eta=\int\limits_0^\infty F^2(r) dr \label{10}
\end{equation}
We have computed the values of $\eta$ and for $m=0$ and two
different values of $\zeta$: $\zeta=0$ and $\zeta=0.3$. The
results are given in Table \ref{table}

\begin{table}
\begin{center}
\caption{\small \(G_A/G_V\) and \(\eta\) for various theoretical
prescriptions in comparison with experimental data}
\begin{tabular}{|c|c|c|c|c|}
\hline
 & Exp. & NRQM &  \(\zeta=0\) &  \(\zeta=0.3\) \\
\hline \(G_A/G_V\) & 1.27 & 1.67 & 1.36 & 1.27 \\ \hline \(\eta\)
& -- & 0 & 0.14 & 0.18 \\ \hline
\end{tabular}
\label{table}
\end{center}
\end{table}

Note that for $m=0$ $G_A$ does not depend on the string tension
$\sigma$ on dimensional grounds. One can see that in the Table
\ref{table} that the resulting $G_A$ is in the correct ballpark
for $\zeta\in [0,0.3]$. The value $\zeta=0.3$ corresponds to the
reasonable effective value of $\alpha_s$ in the $qq$ potential,
namely from
$\left\langle\sum\frac{2}{3}\frac{\alpha_s}{r_{ij}}\right\rangle=
\left\langle\sum\frac{\zeta}{r_i}\right\rangle$, and $\langle
r_{ij}\rangle\approx\sqrt{3}\langle r_i\rangle$, one has
\[
(\alpha_s)_{\text{eff}}=\frac{3\sqrt{3}}{4}\zeta\approx 0.39
\]
and this value of $\zeta$ was checked in the actual calculation of
the nucleon mass \cite{9}. It is rewarding that the resulting
$G_A=1.27$ is in close agreement with experiment.

Concerning the tensor charge $\delta q$, it can be easily
calculated the definition of $\delta q$, and in the same way as in
Eq. (\ref{7}) one arrives at he expression
\begin{equation}
\label{11} \delta q=
\Bigl\langle\chi_\uparrow\Bigm|\beta\bsl{\Sigma}_3\Bigm|\chi_\uparrow\Bigr\rangle
\end{equation}
As a result one has for $\zeta=0.3$:
\begin{equation}
\label{12} \delta q=\delta u-\delta
d=\frac{5}{3}\left(1-\frac{2}{3}\eta\right)\approx 1.47
\end{equation}
Note that in the nonrelativistic limit $\delta q=G_A$.

One can compare these results with the lattice data \cite{10},
where both $G_A$ and $\delta q$ are close to each other and are in
the interval $1.12\le G_A,\delta q\le 1.18$ \cite{10} for
$m_\pi>0.5\text{~GeV}$.

Calculations of $\delta q$ in other methods give results ranging
from 1.07 to 1.45, see \cite{5} for refs. and discussion. Note
that the anomalous dimension of the tensor charge is small and
calculations here and in \cite{5} refer to the scale
$\mu^2=M_N^2$.

There are two possible unaccounted effects which can influence our
results. First, the contribution of other terms in (\ref{4}) --
excited Dirac orbitals. The corresponding multichannel
calculations done in \cite{8} for magnetic moments, result in
decreasing of the modulus of magnetic moments of proton and
neutron by some $10-15\%$ when one accounts for 4 Dirac orbitals
for each quarks, and one can expect the same type of corrections
for $G_A$. Second, the contribution of chiral degrees of freedom,
i.e. of the $\pi$, $\eta$, $K$ exchanges. Again, for nucleon
magnetic moments these corrections are typically of the order of
$10\%$ \cite{8}, and we expect this to be an upper limit for
$G_A$, since magnetic moments are much more sensitive to the
contribution of the lowest Dirac components, than $G_A$ and
$\delta q$, where these contributions enter quadratically and not
linearly.

These corrections are not taken into account above, which is
planned for a subsequent work, where also hyperon semileptonic
decays are considered \cite{11}.

One should note, that relativistic approach to the baryon wave
function, based on the light-cone formalism \cite{12} was shown to
improve the qualitative agreement of $G_A$ with experiment,
however quantitatively still far from experiment.

Summarizing, we have calculated nonsinglet axial and tensor
charges in the simple relativistic model of the nucleon, where the
wave function is a product of three Dirac orbitals of quarks. The
resulting value of $G_A$ is in excellent agreement with experiment
for the choice of the only parameter
$(\alpha_s)_{\text{eff}}=0.39$, yielding the reasonable value of
the nucleon mass. Note that quarks in the baryon prove rather
relativistic, so the contribution of the lower quark bispinor
component to $G_A$ is not negligibly small: $\eta\approx 0.18$.

This work was supported by the federal program of the Russian
Ministry of Industry, Science and Technology \# 40.052.1.1.1112,
by the Grant for support of Leading Scientific Schools \#
1774.2003.2 and in part by the RFBR grant \# 03-02-17345.

\end{document}